\documentclass[10pt, conference]{IEEEtran}

\usepackage{amssymb}
\usepackage{amsmath}

\usepackage[colorlinks=true, linkcolor=blue, citecolor=blue, urlcolor=blue]{hyperref}

\usepackage{graphicx}
\usepackage{booktabs}
\usepackage{multirow}
\usepackage{tabularx}
\usepackage{enumitem}
\usepackage{threeparttable}
\usepackage{balance}
\usepackage{url}

\begin{document}

\title{Snaps: Bloated and Outdated?}

\author{
\IEEEauthorblockN{Jukka Ruohonen}
\IEEEauthorblockA{University of Southern Denmark \\
Email: juk@mmmi.sdu.dk}
\and
\IEEEauthorblockN{Qusai Ramadan}
\IEEEauthorblockA{University of Southern Denmark \\
Email: qura@mmmi.sdu.dk}
}

\maketitle

\begin{abstract}
Snap is an alternative software packaging system developed by Canonical and
provided by default in the Ubuntu Linux distribution. Given the heterogeneity of
various Linux distributions and their various releases, Snap allows an
interoperable delivery of software directly to users. However, concerns and
criticism have also been frequently expressed. Regarding this criticism, the
paper shows that currently distributed snap packages are indeed on average
bloated in terms of their sizes and outdated in terms updating frequencies. With
these empirical observations, this short paper contributes to the research
domain of software packaging, software packages, and package~managers.
\end{abstract}

\begin{IEEEkeywords}
Software packages, package managers, Ubuntu
\end{IEEEkeywords}

\section{Introduction}

Software package distribution in the open source software world has always been
one of the world's advantages but there have always been also challenges. Among
other things, distribution of commercial software---whether for good or
bad---has been challenging due to the world's heterogeneity. To this end, Snap
has provided Linux users with an easy way to install the popular Steam platform
for commercial video games, among other things. However, Snap is not the only
package manager designed to augment the ``native'' package managers in Linux
distributions; Flatpak is another example.\footnote{~\url{https://flatpak.org/}}
These alternatives have faced also frequent criticism. Although snap packages,
or snaps in short, run in sandboxes, security concerns have been raised due to a
suspected lack of updates~\cite{Pope23}, which is paradoxical because an easier
provision of frequent updates was also among the rationales behind
Snap. Analogously to arguments raised in research~\cite{Zhang24}, also excessive
software sizes, or bloat in short, have been a subject of
criticism~\cite{Nardi20}. Like with bloated dependencies~\cite{SotoValero21},
the reason for the alleged bloat is that a snap is a self-contained component,
containing everything required to run a software, including all basic and shared
system libraries used in Linux distributions. With these notes, the following
research questions (RQs) are examined:
\vspace{3pt}
\begin{description}
\itemsep 3pt
\item{$\textmd{RQ}_1$: How large are snaps on average?}
\item{$\textmd{RQ}_2$: How frequently are snaps updated on average?}
\item{$\textmd{RQ}_3$: Do the sizes of snaps explain statistically how
  frequently they are updated?}
\end{description}
\vspace{2pt}


\section{Data}

The dataset, which is available online \cite{Ruohonen25Data}, was assembled by
retrieving all snaps available for a x86\_64 installation of Ubuntu Desktop
24.04.02 LTS. The list of available snaps is provided as a file in this Ubuntu
release.\footnote{~\texttt{/var/cache/snapd/names}} A separate online repository
is also available.\footnote{~\url{https://snapcraft.io/}} At the start of the
data collection on 20 June 2025, there were $6,908$ snaps listed in the file. Of
these, $n = 6,797$ were successfully retrieved from the repository via the
\texttt{snap} command with the option \texttt{download}. Most of the failed
cases involved snaps that were no longer available in the repository. A few
other cases also failed during extraction using the \texttt{unsquashfs} command
because these required root privileges to create character devices. Then, the
following variables were constructed:
\begin{itemize}
\itemsep 3pt
\item{\textit{LastUpdated} refers to the time differences between the
  downloading dates and the latest stable releases available for the snaps. The
  latter was obtained using the \texttt{snap} command with the \texttt{info}
  option, and the character string \texttt{latest/stable} was used to extract
  the release dates.}
\item{\textit{Megabytes} refers to the size of a fully extracted snap in
  megabytes, as given by the standard \texttt{du} command.}
\item{\textit{Files} refers to the number of regular files present in an
  extracted snap, excluding symbolic links and others.}
\item{\textit{Binaries}, \textit{Scripts}, and \textit{Libraries} refer to the
  number of binaries, shell scripts, and shared libraries present in the snaps,
  respectively. These were identified using the standard \texttt{file} command
  by using the character strings \texttt{LSB executable}, \texttt{ASCII text
    executable}, and \texttt{LSB shared object} for the identification,
  respectively.}
\end{itemize}

All variables except \texttt{LastUpdated} measure software size. However---and
interestingly enough, they are not all highly correlated with each other. The
maximum (Pearson) correlation coefficient is $0.73$ between \textit{Files} and
\textit{Libraries}. The remaining coefficients are all below $0.40$ in absolute
value.

\section{Methods}\label{sec: methods}

Descriptive statistics are sufficient for answering to the research questions
$\textmd{RQ}_1$ and $\textmd{RQ}_2$. To examine $\textmd{RQ}_3$, the following
five separate Poisson regression models are estimated:
\begin{equation*}
\begin{cases}
\ln(\textmd{E}[\textit{LastUpdated}~\vert~\textit{Megabytes}]) = \alpha_1 + \beta_1\textit{Megabytes} , \\ \notag
\ln(\textmd{E}[\textit{LastUpdated}~\vert~\textit{Files}]) = \alpha_2 + \beta_2\textit{Files} , \\ \notag
\ln(\textmd{E}[\textit{LastUpdated}~\vert~\textit{Binaries}]) = \alpha_3 + \beta_3\textit{Binaries} , \\ \notag
\ln(\textmd{E}[\textit{LastUpdated}~\vert~\textit{Scripts}]) = \alpha_4 + \beta_4\textit{Scripts} , \\ \notag
\ln(\textmd{E}[\textit{LastUpdated}~\vert~\textit{Libraries}]) = \alpha_5 + \beta_5\textit{Libraries} .
\end{cases}
\end{equation*}

Above, $\ln(\cdot)$ denotes the natural logarithm and $\textmd{E}(\cdot)$ the
expected value, $\alpha_1, \ldots, \alpha_5$ are constants, and $\beta_1,
\ldots, \beta_5$ are regression coefficients. Then, the verb ``explain'' used in
$\textmd{RQ}_3$ requires decision-making material. Two criteria are used.

The first criterion is that all estimated regression coefficients,
$\hat{\beta}_1, \ldots, \hat{\beta}_5$, must be statistically significant at the
conventional 95\% level. The second criterion is that at least one coefficient
must show a visible effect. To subsequently define what the adjective
``visible'' means, a percentage increase in $\textit{LastUpdated}$ for a unit
increase in any of the independent variables seems like a reasonable threshold
for a minimum visible effect. Such an effect is small but still likely not mere
noise. For a positive answer to $\textmd{RQ}_3$, there must thus also exist a
$\hat{\beta}_i$, given $i = 1, \ldots, 5$, which satisfies $\exp(\hat{\beta}_i)
- 1 \geq 1$.

\section{Results}

\begin{figure}[th!b]
\centering
\includegraphics[width=\linewidth, height=4.2cm]{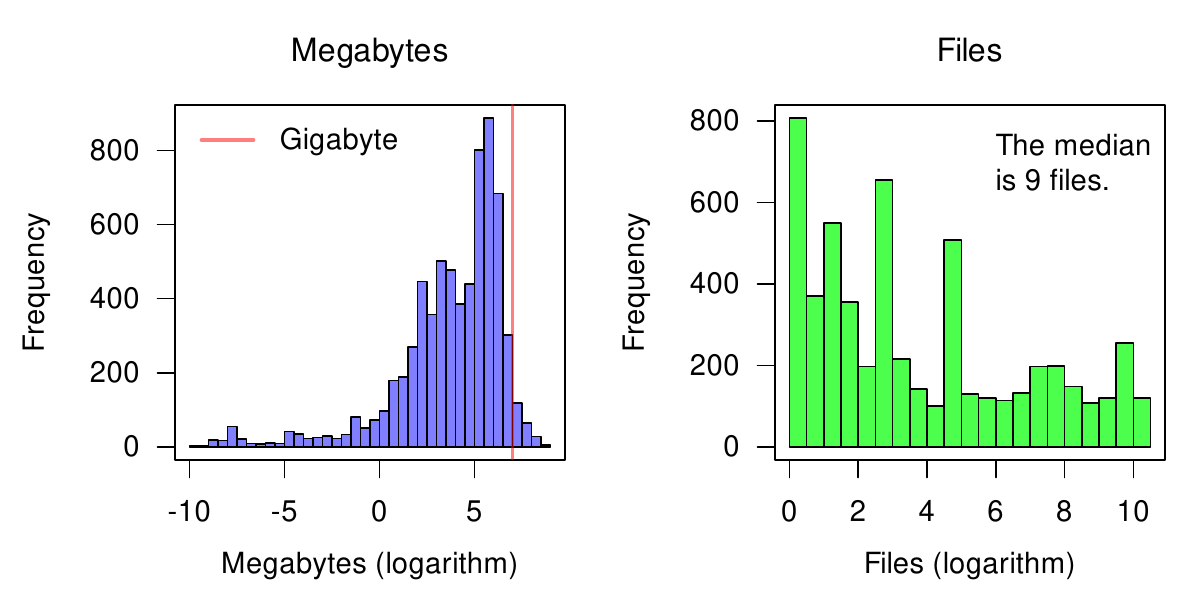}
\caption{The Package Sizes}
\label{fig: sizes}
\end{figure}

The results can be disseminated by going through the three research questions
consecutively. Thus, Fig.~\ref{fig: sizes} shows the histograms of the natural
logarithms of \textit{Megabytes} and \textit{Files}. Both indicate long-tailed
distributions, as is typical for software size variables~\cite{Lindsay10}. The
means and medians are $235$ and $83$ megabytes, respectively, and $9$ and $1610$
files. These numbers allow to answer to $\textmd{RQ}_1$ by saying that on
average snaps are large enough to be characterized as bloated. However, it is
worth noting that the snaps are still less bloated than popular machine learning
containers from which about a half has been observed to be larger than $10$
gigabytes~\cite{Zhang24}. Also the \textit{Binaries}, \textit{Scripts}, and
\textit{Libraries} variables display almost identically shaped long-tailed
empirical probability distributions, although the averages are obviously much
lower.

\begin{figure}[th!b]
\centering
\includegraphics[width=\linewidth, height=3.7cm]{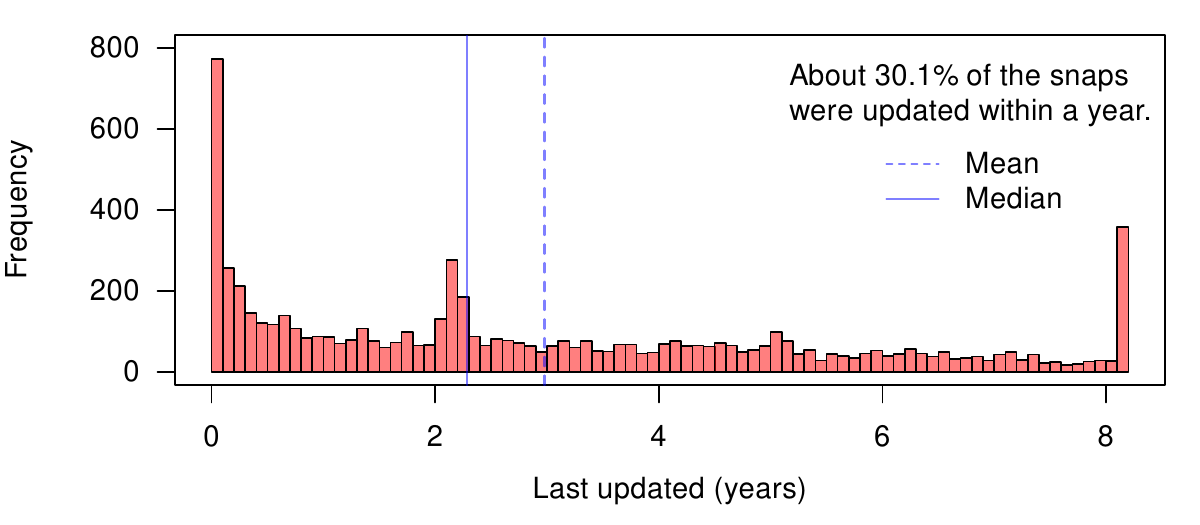}
\caption{The Updating Pace}
\label{fig: update}
\end{figure}

Regarding $\textmd{RQ}_2$, a histogram of the \textit{LastUpdated} variable is
shown in Fig.~\ref{fig: update}. Unlike with many similarly constructed time
difference variables~\cite{Ruohonen25SANER}, the distribution's right tail is
long but not rapidly decreasing after the peaking mass at the left tail. Both
the mean and median are more than two but less than four years. Again, these
averages are enough to answer $\textmd{RQ}_2$ by saying that snaps are
infrequently updated on average. Many have not seen updates in the last four
years. There are a few outliers that have not been updated even in over eight
years.

Finally, the answer to $\textmd{RQ}_3$ is that the snaps' sizes do not explain,
or explain only poorly, the frequency of updates. Although all regression
coefficients of the independent variables are statistically significant at the
threshold chosen, their magnitudes are small: $\vert\hat{\beta}_1\vert$,
$\vert\hat{\beta}_2\vert$, $\vert\hat{\beta}_4\vert$, and
$\vert\hat{\beta}_5\vert$ are all below a value $0.0005$. Even the largest
effect, $\hat{\beta}_3 = 0.0106$ for \textit{Binaries}, is below the minimum
threshold that was specified~in~Section~\ref{sec: methods}.

\section{Conclusion}

The conclusion does not comply with the so-called Betteridge's law of headlines
according to which a title that ends in a question mark entails a negative
answer. In other words, the conclusion is that snaps indeed are bloated
($\sim\textmd{RQ}_1$) and outdated ($\sim\textmd{RQ}_2$) on average. However,
the sizes of the nearly seven thousand snaps observed do not explain well the
updating lags~($\sim\textmd{RQ}_3$). What about practical implications?

Regarding users---and even though security was not a part of the RQs, it seems
sensible to recommend care when using snaps because some of these presumably
contain identified and disclosed vulnerabilities. This recommendation does not
mean that all snaps would be insecure and unsafe to use. For instance, the Snap
packages for the Chromium and Firefox web browsers are frequently updated. With
respect to Ubuntu and Canonical, the existing email notification system to
prompt developers about updating outdated snaps~\cite{SotoValero21} does not
seem to be functioning well. Thus, better monitoring and a more aggressive
deprecation policy might be something to consider.

\bibliographystyle{abbrv}

\begin{thebibliography}{1}

\bibitem{Lindsay10}
J.~Lindsay, J.~Noble, and E.~Tempero.
\newblock {D}oes {S}ize {M}atter? {A} {P}reliminary {I}nvestigation of the
  {C}onsequences of {P}owerlaws in {S}oftware.
\newblock In {\em Proceedings of the 2010 ICSE Workshop on Emerging Trends in
  Software Metrics (WETSoM 2010)}, pages 16--23, Cape Town, 2010. ACM.

\bibitem{Nardi20}
T.~Nardi.
\newblock {W}hat's the {D}eal {W}ith {S}nap {P}ackages?
\newblock Available onine in June 2025:
  \url{https://hackaday.com/2020/06/24/whats-the-deal-with-snap-packages/},
  2020.

\bibitem{Pope23}
A.~Pope.
\newblock {O}utdated {S}nap {P}ackages.
\newblock Available online in June 2025:
  \url{https://blog.popey.com/2023/09/outdated-snap-packages/}, 2023.

\bibitem{Ruohonen25Data}
J.~Ruohonen.
\newblock {A} {D}ataset for a {P}aper {E}ntitled ``{S}naps: {B}loated and
  {O}utdated?''.
\newblock Zenodo, available online:
  \url{https://doi.org/10.5281/zenodo.15774457}, 2025.

\bibitem{Ruohonen25SANER}
J.~Ruohonen.
\newblock {A}n {E}mpirical {S}tudy of {V}ulnerability {H}andling {T}imes in
  {CP}ython.
\newblock In {\em Proceedings of the IEEE International Conference on Software
  Analysis, Evolution and Reengineering (SANER 2025)}, pages 891--896,
  Montreal, 2025. IEEE.

\bibitem{SotoValero21}
C.~Soto-Valero, N.~Harrand, M.~Monperrus, and B.~Baudry.
\newblock {A} {C}omprehensive {S}tudy of {B}loated {D}ependencies in the
  {M}aven {E}cosystem.
\newblock {\em Empirical Software Engineering}, 26:1--44, 2021.

\bibitem{Zhang24}
H.~Zhang, M.~Alhanahnah, F.~A. Ahmed, D.~Fatih, P.~Leitner, and A.~Ali-Eldin.
\newblock {M}achine {L}earning {S}ystems {A}re {B}loated and {V}ulnerable.
\newblock {\em Proceedings of the ACM on Measurement and Analysis of Computing
  Systems}, 8(1):1--30, 2024.

\end{thebibliography}

\end{document}